\documentclass[aps,prb,preprint,showpacs]{revtex4-1}
\usepackage{graphicx}
\begin{document}
\title{Hidden-anisotropy-induced $\pi$ phase shift in all-optical magnetization precession}
\author{D. Wang}
\email{daowei.wang@matfyz.cuni.cz}
\affiliation{Department of Condensed Matter Physics, Charles University, 12116 Prague, Czech Republic}
\date{\today}
\begin{abstract}
Laser-induced magnetization precession of an in-plane magnetized Pt/Co/Pt film with perpendicular interface anisotropy was studied using time resolved magneto-optical Kerr effect. An additional $\pi$ shift in the phase of precession is needed to describe the measured data if only the demagnetization energy is considered. Based on an augmented microscopic model description of the laser-induced magnetization dynamics, the additional $\pi$ phase is found to be rendered by the dependence on the phonon temperature of the hidden interface anisotropy, in contrast to the dependence on the electron temperature of the demagnetization energy. The observation that the phase of precession is affected by both the electron and the phonon temperature warrants a detailed knowledge about the forms of anisotropy present in the system under investigation for a holistic description of laser-induced magnetization precession.
\end{abstract}
\maketitle
\section{Introduction}
\label{intro}
Since the first experimental demonstration of ultrafast demagnetization in  ferromagnetic Ni in 1996 \cite{beaurepaire96}, the interplay between coherent light and magnetic order has attracted much attention in the magnetism community \cite{Kirilyuk10}. The physics involved in the ultrafast demagnetization is so complicated that, almost 30 years after its discovery, the microscopic mechanism responsible for the transfer of angular momentum between electron, spin and lattice subsystems, upon irradiation by laser pulses, remains elusive. Possible candidates include direct angular momentum transfer from photons to electrons \cite{Zhang00}, electron-phonon scattering \cite{koopmans05,koopmans10,Griepe23}, electron-magnon scattering \cite{Carpene08}, electron-electron scattering \cite{Krauss09}, and coherent interaction between electrons and photons \cite{bigot09}. In contrast to these local dissipation channels, superdiffusive transport due to the different lifetime for spin-up and spin-down electrons was proposed to account for the demagnetization observed in the first several hundred femtoseconds after laser irradiation \cite{battiato10,battiato12}. For a complete description of the ultrafast demagnetization in ferromagnets, all of those processes should be included in a Boltzmann-like approach \cite{Mueller11}, with information provided by complementary experimental techniques that probe separately the magnetization \cite{Kirilyuk10}, electron \cite{Cinchetti06,Eich17}, and phonon \cite{Zahn21} dynamics.

A related phenomenon occurring on a longer timescale is the laser-induced magnetization precession in ferromagnetic metals \cite{van Kampen02prl,van Kampen02jmmm}. Depending on the anisotropy of the studied material, the precession period can vary drastically. But the typical timescale is of the order of 0.1 ns. The magnetization precession observed can be understood on the basis of a change in anisotropy, which is a sensitive function of temperature. Intuitively, the two processes, i.e. the ultrafast demagnetization occurring on the timescale of 0.1 ps and the magnetization precession with periods of about 0.1 ns, are connected to each other. Actually, with a two temperature model (2TM) \cite{bigot05}, the magnetization precession was explained as a consequence of the dynamic temperature profile, which is just the driving force for the ultrafast demagnetization \cite{koopmans10}.
\begin{figure}\centering
\begin{minipage}[c]{0.5\linewidth}
\includegraphics[width=\linewidth]{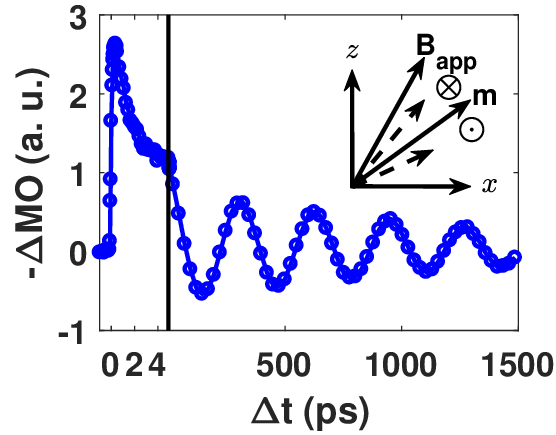}
\end{minipage}
\caption{Representative laser-induced magnetization dynamics for a 10 nm permalloy (Ni$_{81}$Fe$_{19}$) film with an external field $B_{app}$ = 0.15 T applied almost perpendicular to the film plane \cite{DallaLonga08}. The black vertical line highlights $\Delta t$ = 5 ps, where the change of time scale occurs. Dashed arrows in the inset give schematically the effective fields for initial magnetization precession after laser excitation: the dashed arrow together with the direction of its precessional torque, shown above the magnetization, corresponds to the case for permalloy, i.e. only demagnetization anisotropy is present, while the dashed arrow and torque direction below \textbf{m} corresponds to the presence of both demagnetization and interface anisotropy, which is appropriate for Pt/Co/Pt films. Plot of the inset is schematic and not to scale.}
\label{py}
\end{figure}

For a thin film of metallic ferromagnetic material under the influence of an out-of-plane field, if there is no other forms of anisotropy present except for the shape, or demagnetization, anisotropy, the effective demagnetization field decreases in magnitude for the first several hundred femtoseconds, following the ultrafast demagnetization process caused by laser heating. As a result, the total effective field is further tilted out of plane, and the magnetization vector will precess instantaneously around the new effective field. Hence, in this case, the initial precession of the magnetization is towards the direction of the external field (cf. the inset to Fig. \ref{py}), corresponding to an initial phase of $-\pi/2$. This typical behaviour is routinely observed in magnetic films with easy-plane anisotropy, an example of which is shown in Fig. \ref{py} \cite{DallaLonga08}. Atomistic simulation based on the Landau-Lifshitz-Bloch (LLB) equation \cite{Kazantseva08} also supports this picture \cite{Atxitia07}.

\begin{figure}\centering
\begin{minipage}[c]{0.8\linewidth}
\includegraphics[width=\linewidth]{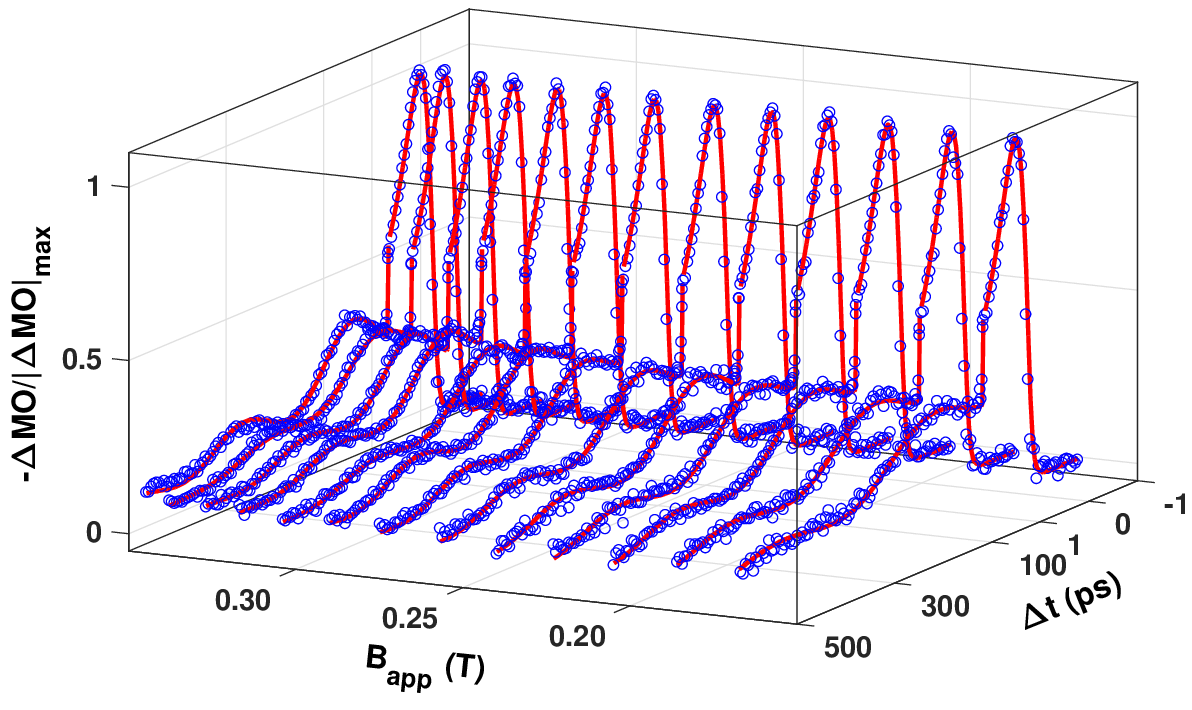}
\end{minipage}
\caption{Magneto optical traces with various applied field ($B_{app}$) normalized to the maximal signal, $|\Delta MO|_{max}$. Note the change of scale at delay time $\Delta t$ = 1 ps. Blue circles are experimental data, and red lines are the corresponding fits. Distinctive features in magnetization dynamics induced by ultrashort laser pulses irradiating on ferromagnetic metals are discernible: ultrafast demagnetization occurring on the timescale of $<$ 1 ps and magnetization precession on the order of $\sim$ 0.1 ns.}
\label{mosig}
\end{figure}

However, for in-plane magnetized Pt/Co/Pt films, our measurements give a completely different behaviour: the magnetization initially moves away from the direction of the external field, as plotted in Fig. \ref{mosig}, although static hysteresis loops determine unambiguously that the film plane is an easy-plane. This motion of magnetization is described by an initial phase of $\pi/2$, bringing about a phase difference of $\pi$. A similar difference in the initial precession was observed in Ref. \onlinecite{bigot05} for thick Co films deposited on Al$_2$O$_3$ and MgO substrates. Due to the different preferred growth orientations of the $c$ axis of hexagonal Co, the axis of uniaxial crystalline anisotropy is in-plane on MgO but perpendicular-to-plane on Al$_2$O$_3$. The observed difference in initial precession was thus attributed to the different orientations of the uniaxial anisotropy. Theoretical simulation qualitatively explained the difference observed, based on a molecular-field model description for the ultrafast demagnetization and a Bloembergen relaxation-time approximation for the magnetization dissipation, but no direct comparison between experiment and simulation was performed and no quantitative conclusion reached. If we still want to stick to the picture that the magnetization precession is initiated by the ultrafast demagnetization, which also serves as a test of our microscopic description of the ultrafast demagnetization, the observed $\pi$ disparity in phase has to be resolved and a quantitative agreement between experiment and theory achieved, which is the main motivation for the current work.

Using a microscopic model description for the longitudinal relaxation of magnetization and the Landau-Lifshitz-Gilbert (LLG) equation for the transverse relaxation, we achieve a reasonable quantitative agreement between experiment and theory for the description of the magnetization dynamics triggered by laser heating with feasible parameters, and find that the discrepancy in the initial phase can be removed by considering the hidden interface anisotropy's different temperature dependency as compared to that of the demagnetization anisotropy. The interface anisotropy is termed as hidden simply because its existence is not felt at all in equilibrium; an effective demagnetization anisotropy, which includes the interface anisotropy, is enough to describe the static behavior of our in-plane magnetized films. The fact that the interface anisotropy is hidden behind the effective demagnetization anisotropy highlights an important difference from Ref. \onlinecite{bigot05}: the $\pi$ phase shift observed here is not caused by a change for the effective anisotropy from being in-plane to perpendicular-to-plane, but by a local competition between the perpendicular-to-plane interface anisotropy and the in-plane demagnetization anisotropy; the effective anisotropy is always in-plane for the whole film, and the additional $\pi$ phase needed to describe the experimental data is relative to the effective in-plane anisotropy. As a result of the different temperature dependencies, the needed $\pi$ shift in phase for magnetization precession follows naturally, in consistence with both Figs. \ref{py} and \ref{mosig}. The model analysis shows that the additional phase is mainly determined by the time evolution of the total effective field and accumulated in the first several picoseconds, during which period the magnetization precesses around the time-varying effective field. Consequently, the accumulated phase depends on the history of the effective field and a definite phase of precession is only meaningful after a specific time delay from the arrival of the laser pulse, which is in stark contrast with the conventional description of the laser-induced magnetization precession. There the change in anisotropy is instantaneous and, hence, the phase of precession is defined at delay time zero. Another quantitative discrepancy with the conventional, qualitative description for the initial magnetization precession is that the actual phase differs from the exact values $\pm \pi/2$ by a finite amount. When those two quantitative discrepancies are taken into account, surprisingly, the definite $\pi$ difference is recovered by considering the temperature evolution of the hidden interface anisotropy. Therefore, our analysis for the role played by the hidden interface anisotropy in determining the initial phase is a step further towards the quantitative understanding on the microscopic origin of the laser-induced magnetization precession, and can provide conceptual insight into the mechanism behind magnetization precession either in more complicated material systems \cite{Mizukami15,Bonda20} or affected by other forms of anisotropy, such as magneto-elastic interaction \cite{Kats16,Jarecki24}.

The organization of the article is as follows. After this Introduction, in Sec. \ref{model} we will briefly discuss the augmented microscopic model for the description of the laser-induced magnetization dynamics. Sec. \ref{results} will then give the experimental results and the numerical fitting of the experimental data to the theoretical model. The resultant $\pi$ phase shift will be determined there, after a detailed analysis. Our conclusion of the current investigation will be given in Sec. \ref{conclusion}. For comparison, an Appendix is given finally to show the behaviour of the precession frequency and the decay time, in addition to the precession phase, which is the focus of the current study.

\section{Theoretical description of magnetization dynamics}
\label{model}
In our model description, the ultrafast demagnetization is described by the microscopic three temperature model (M3TM) \cite{koopmans10}, and the transverse relaxation of magnetization is given by the phenomenological LLG equation \cite{llg1,llg2}. In spirit, the separation of the magnetization dynamics into longitudinal and transverse relaxations used here is similarly employed in the LLB equation \cite{Kazantseva08} and the self-consistent Bloch equation \cite{Xu12}. The only difference lies in the longitudinal relaxation term, which is given here by the M3TM \cite{koopmans10}. In the M3TM, if only heat dissipation along the film thickness is considered, the time evolution of the electron temperature $T_e$ and the phonon temperature $T_p$ is determined essentially by the 2TM,
\begin{eqnarray}
C_e \frac{d T_e}{dt} &=& \nabla_z (\kappa \nabla_z T_e) + g_{ep} (T_p - T_e) + P(t), \nonumber\\
C_p \frac{d T_p}{dt} &=& g_{ep} (T_e - T_p),
\label{ttm}
\end{eqnarray}
where $C_e$ and $C_p$ are the corresponding heat capacities. $P(t)$ is the laser power density absorbed by the electron subsystem, and $\nabla_z$ denotes the $z$ component of the gradient operator. $\kappa$ is the electronic thermal conductivity of Co, and $g_{ep}$ the electron-phonon coupling constant. $g_{ep}$ is assumed to be a constant, although it is actually a temperature dependent quantity \cite{Lin08}. Microscopically, $g_{ep} = 3 \pi D_F ^2 D _p E _D k _B \lambda _{ep} ^2/ 2 V _{at} \hbar$, where $\hbar$ is the reduced Planck's constant, $D_p$ the number of atoms per atomic volume $V_{at}$, $E_D$ the Debye energy, $D_F$ the electronic density of states at the Fermi energy, $k_B$ the Boltzmann constant, and $\lambda_{ep}$ the microscopic electron-phonon coupling constant.

The dynamics of \textbf{m} = \textbf{M}/$M_0$, the magnetization vector normalized to saturation magnetization $M_0$ at zero temperature, is governed by
\begin{equation}
\frac{d\textbf{m}}{dt} = R T_p \left[ 1 - m \coth \left(\frac{m T_C}{T_e} \right)\right] \textbf{m} - \gamma \left[ \textbf{m} \times \textbf{B} +  \frac{\alpha}{m} \textbf{m} \times (\textbf{m} \times \textbf{B})\right]
\label{mllg}
\end{equation}
with $m$ being the magnitude of \textbf{m}, $\gamma$ the gyromagnetic ratio, $T_C$ the Curie temperature, $\alpha$ the Gilbert damping constant, and \textbf{B} the total effective magnetic field, including the external, anisotropy and demagnetizaion field contributions. Constant $R$ determines the demagnetization rate, and is related to the spin-flip probability $\alpha_{sf}$ during electron-phonon collisions, mediated by the spin-orbit coupling, through $R = 8 \alpha_{sf} g_{ep} k_B T_C / E_D ^2 \mu_{at}$ with $\mu_{at}$ the number density of Bohr magnetons. The main modification made here to the conventional M3TM \cite{koopmans10} is the addition of the transverse relaxation term in Eq. (\ref{mllg}).

\begin{table}[htb]
\begin{tabular}{|c|c|c|c|c|c|c|}
\hline
$A$(pJ/m)&$C_T$(MJ/m$^3$K)&$E_D$ (meV)&$\kappa$ (W/mK)&$M_0$($\mu_B/V _{at}$)&$T_C$(K)&$V _{at}$({\AA}$^3$)\\
\hline
28 \cite{Liu96,Grimsditch97}&3.73 \cite{koopmans10}&38.4 \cite{kittel05}&40 \cite{Dejene12}&1.72 \cite{Stohr06}&1388 \cite{Stohr06}&11.1 \cite{Stohr06}\\
\hline
\end{tabular}
\caption{Material parameters used in our theoretical description of the magnetization dynamics induced by laser heating, which are fixed to the values given in this table in the fitting to the experimental data.}
\label{paras}
\end{table}

It is well known that, at Pt/Co interfaces, the interface anisotropy is perpendicular to the film plane, due to the 3\emph{d}-5\emph{d} hybridization there \cite{Bruno89,Stohr99}. Assuming negligible bulk anisotropy, the total anisotropy field is correspondingly comprised of the interface anisotropy field
\begin{equation}
B_s \frac{m^3(T_p)} {m ^2} m_z \hat{\textbf{e}}_z [\delta (z) + \delta (z - z_0)],
\label{bk}
\end{equation}
and the demagnetization field $- \mu_0 M_0 m _z \hat{\textbf{e}}_z$. $B_s = 2 K_s/M_0$ is the zero-temperature interface anisotropy field, with $K_s$ the corresponding anisotropy constant. $\hat{\textbf{e}}_z$ is the unit vector perpendicular to the film plane. $z = 0$ and $z = z _0$ correspond to the two Co/Pt interfaces. The temperature dependence of the interface anisotropy is given explicitly in Eq. (\ref{bk}) by the term cubic \cite{Callen65,kisielewski12} in $m$. Note that we have postulated that the interface anisotropy is sensitive to the lattice temperature $T_p$, as it is primarily determined by the crystal field \cite{bigot05}. Due to the interface character of $B_s$, there is a critical Co thickness where transition from out-of-plane to in-plane magnetized configuration occurs, which is around 1 nm for our sputtered samples \cite{Lavrijsen11}. Hence for the 4 nm Co film considered here, the demagnetization field dominates and the film is magnetized in-plane at remanence.
\begin{figure}\centering
\begin{minipage}[c]{0.5\linewidth}
\includegraphics[width=\linewidth]{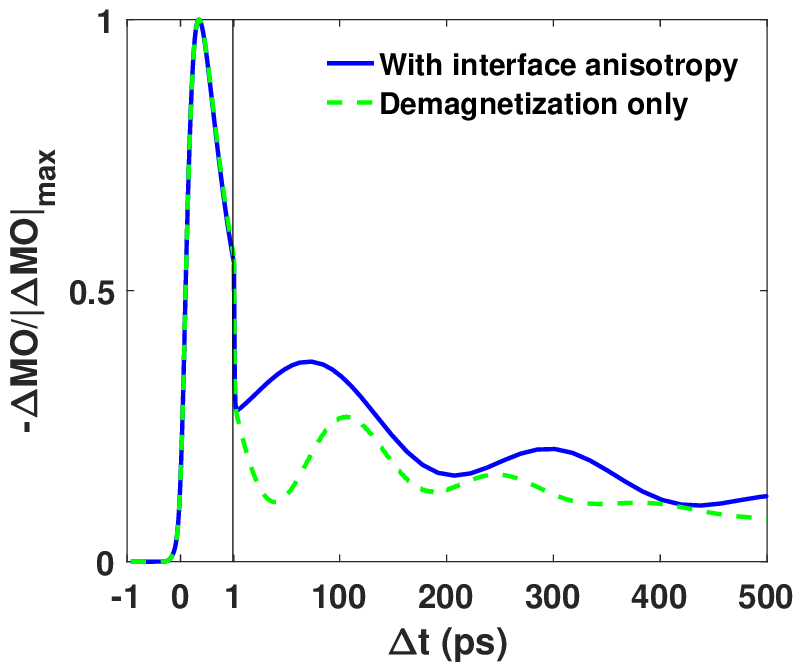}
\end{minipage}
\caption{Simulated magneto-optical traces with contribution from only $m _z$ at $B _{app}$ = 0.35 T in the presence of the interface anisotropy ($K _s \neq 0$) or not ($K _s = 0$). The transition of time scale at $\Delta t$ = 1 ps is marked by the vertical black line.}
\label{pma}
\end{figure}

In order to numerically study the laser induced magnetization dynamics, the Co film was divided into four layers. The top layer and the bottom layer are affected by both the interface anisotropy field and the demagnetization field, while the middle layers are only influenced by the demagnetization field. The exchange coupling between adjacent layers $i$ and $j$ is modelled by the usual expression $E_X = - A \textbf{m}_i \cdot \textbf{m}_j/d ^2$, with $d$ = 1 nm being the separation between adjacent layers and $A$ the exchange stiffness constant. The laser pulse $P(t)$ was modeled by a gaussian function with group velocity dispersion \cite{Diels06}. The optical penetration depth at 780 nm of Co is 13.5 nm \cite{Krinchik68}. Except for the magnitude of the laser pulse, all other optical parameters used in the simulation were extracted from numerically fitting the short timescale ($<$1 ps) demagnetization data, using a phenomenological model given in Ref. \onlinecite{Malinowski08}. Material parameters employed in the model description of magnetization dynamics using Eq. (\ref{mllg}) are given in Table \ref{paras}, which are all bulk Co values except for $\kappa$ and $A$. The heat exchange between the Co film and the substrate is treated simply by a phenomenological thermal conductivity $\kappa_{sub}$, which is varied to fit the measured data, as well as the ratio \cite{koopmans10} $C _e/ C _p$. The substrate temperature is set to the ambient temperature, $T_{am}$ = 300 K.
\begin{figure}\centering
\begin{minipage}[c]{0.8\linewidth}
\includegraphics[width=\linewidth]{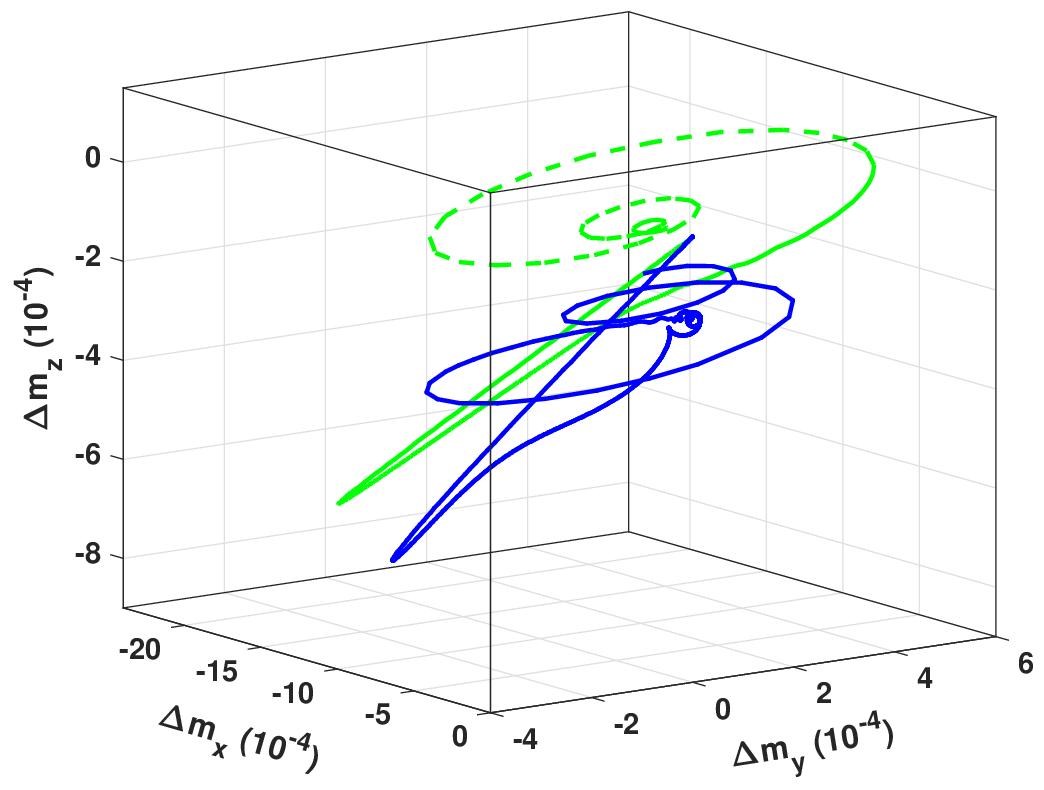}
\end{minipage}
\caption{Simulated magnetization trajectories for the top Co layer at $B _{app}$ = 0.29 T in the presence of the interface anisotropy ($K _s \neq 0$, solid blue curve) or not ($K _s = 0$, dashed green curve). For clarity, the trajectory with $K _s = 0$ is scaled up by a factor of two along the $z$ direction.}
\label{dm}
\end{figure}

\section{Experimental and numerical results}
\label{results}
The sample investigated was a Pt (4 nm)/Co (4 nm)/Pt (2 nm) film made by DC magnetron sputtering onto a Boron doped Silicon wafer with 100 nm thermally oxidized SiO$_2$. The base pressure of the sputtering chamber was 5.0 $\times$ 10$^{-8}$ mbar. The Ar sputtering pressure for Pt was 3.0 $\times$ 10$^{-3}$ mbar, while it was 1.0 $\times$ 10$^{-2}$ mbar for Co. The sputtering rate was 1.16 {\AA}/s for Pt and 0.29 {\AA}/s for Co. Time-resolved magneto optical Kerr effect (TRMOKE) measurements were performed using a pulsed Ti:Sapphire laser with central wavelength 780 nm, pulse width 70 fs and repetition rate 80 MHz. Both pump and probe beams were focused onto the sample at almost normal incidence, hence the measured TRMOKE signal is most sensitive to the out-of-plane ($z$) component of the magnetization. The laser pump pulses induced, delay time ($\Delta t$) dependent Kerr rotation was recorded using a double modulation technique \cite{koopmans00}. In the TRMOKE measurements, the external magnetic field was applied almost normal to the film ($xy$) plane, in order to tilt the magnetization out of the film plane.

Experimental TRMOKE traces and best fits are shown in Fig. \ref{mosig}, after subtracting the state filling effect contribution \cite{koopmans00prl} at $\Delta t$ = 0 to the experimental data. In fitting to the experimental data, the measured magneto-optical signal, $MO$, is assumed to have contributions from all three components of the magnetization vector \cite{Yang93,Qiu00}, $MO \propto m_z + \alpha_x m_x + \alpha_y m_y$, given that, in our experimental setup, the magnetization has not only the $z$ but all three components. Then the variation of the magneto-optical signal, $\Delta MO$, which is defined as the difference after and before the arrival of the laser pulse, is normalized to the maximal demagnetization, $|\Delta MO|_{max}$, as shown in Fig. \ref{mosig}. The normalized data is then fitted by Eq. (\ref{mllg}) with $T _e$ and $T _p$ determined by Eq. (\ref{ttm}). Details of the fitting procedure can be found in Ref. \onlinecite{koopmans10}. It can be seen from Fig. \ref{mosig} that the overall agreement between experiment and theory is satisfactory, considering the crudeness of our model. The agreement shows that the main physics is capture by the simple Eqs. (\ref{ttm}) and (\ref{mllg}). The relevant physical parameters obtained from the best fits are given in Table \ref{fitparas}, where the errors given are the standard deviations of fitted values corresponding to different applied field $B_{app}$. The fitted $\gamma$ corresponds to a Land\'{e} g-factor $g$ = 1.86 $\pm$ 0.07, which is very close to the free electron value. The interface anisotropy gives an out-of-plane to in-plane transition thickness around 2.3 nm at zero temperature. This value is two times of the experimental value of about 1 nm. Since we used the bulk $T_C$ and $\mu_{at}$ in the fitting procedure, this difference is still acceptable. Finally, the Elliott-Yafet spin-flip probability $\alpha_{sf}$ and the electron-phonon coupling constant $\lambda_{ep}$ are comparable to those obtained in Ref. \onlinecite{koopmans10}.
\begin{table}[htb]
\begin{tabular}{|c|c|c|c|c|}
\hline
$\alpha$($10^{-2}$)&$\alpha_{sf}$($10^{-1}$)&$\gamma$($10^{11}$Hz/T)&$K_s$(mJ/m$^2$)&$\lambda_{ep}$(meV)\\
\hline
7 $\pm$ 2&1.6 $\pm$ 0.1&1.64 $\pm$ 0.06&1.50 $\pm$ 0.01&11.3 $\pm$ 0.3\\
\hline
\end{tabular}
\caption{Material parameters determined from fitting the experimental data to the theoretical equations (\ref{ttm}) and (\ref{mllg}).}
\label{fitparas}
\end{table}

To confirm that the observed difference in the initial magnetization precession is actually caused by the interface anisotropy, we can set $K _s$ to zero and keep other parameters intact, thus eliminating the effect of the interface anisotropy and retaining only the demagnetization field. The result for $B _{app}$ = 0.29 T is shown in Fig. \ref{pma}, which demonstrates clearly the effect of the interface anisotropy and that the different behaviours for the initial magnetization precession are in qualitative agreement with what we can expect. To see more clearly the magnetization dynamics and the difference in the initial precession, we plot in Fig. \ref{dm} the trajectories calculated for the top Co layer using the same set of parameters with or without the interface anisotropy. It can be easily seen that, in the presence of only the demagnetization field, the magnetization vector keeps moving upward after the ultrafast demagnetization and recovery process, continuing the trend of the magnetization recovery; while if the hidden interface anisotropy is present, the magnetization's initial precession is downward and against the tendency of the magnetization recovery. The net effect of the competition between the two forms of anisotropy is an almost $\pi$ change in the phase of magnetization precession.

Another interesting feature shared by both trajectories shown in Fig. \ref{dm} is the development of the positive $y$ component of the magnetization vector during the ultrafast demagnetization process. Usually, it is assumed that the ultrafast demagnetization can influence the magnetization procession following it by setting the precession's initial status, but the effect of the transverse magnetization relaxation on the ultrafast demagnetization is negligible. This is true if only demagnetization and remagnetization times are concerned, as they are mainly determined by $\alpha _{sf}$ and $\lambda _{ep}$. However, as demonstrated clearly by the development of magnetization component perpendicular to the initial $xz$ plane in Fig. \ref{dm}, the actual magnetization motion during the ultrafast demagnetization process is indeed modified by the transverse magnetization relaxation; the magnetization vector will remain completely in the $xz$ plane if the transverse relaxation term is absent in Eq. (\ref{mllg}), which is the case in the conventional description of the ultrafast demagnetization.
\begin{figure}\centering
\begin{minipage}[c]{0.5\linewidth}
\includegraphics[width=\linewidth]{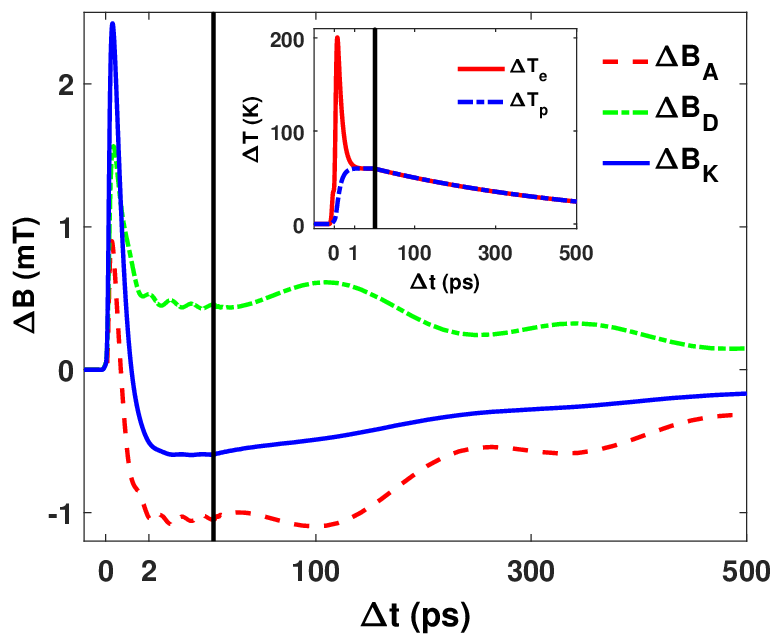}
\end{minipage}
\caption{Dynamic evolution of the interface anisotropy field ($B_A$, dashed line), the demagnetization field ($B_D$, dash-dotted line) and the total effective field ($B_K$, solid line) for the top Co layer with $B_{app}$ = 0.29 T. The inset shows correspondingly the change of the electron temperature and the lattice temperature with delay time $\Delta t$. Note the change of scale at $\Delta t$ = 5 ps (2 ps in the inset) delineated by the vertical solid line. The small amplitude ringing structure visible in $\Delta B_D$ and $\Delta B_A$, which attenuates to zero in about 20 ps, is caused by the interlayer exchange coupling.}
\label{hk}
\end{figure}

Having established that taking into account of the hidden interface anisotropy can qualitatively reproduce the observed phase shift in the initial precession, it is desirable to unfold what actually happens behind by following the time evolution of the total anisotropy field $\Delta B_K = \Delta B _A + \Delta B _D$ for the top Co layer as plotted in Fig. \ref{hk}, together with its two competing components, the interface anisotropy field $B _A = B _s m _z m ^3 (T _p)/m ^2$ and the demagnetization field $B_D = - \mu _0 M _0 m _z$. The main characteristics of Fig. \ref{hk} is that, while the short timescale variation of $B_D$ and $B_A$ is both positive, $\Delta B_D$ and $\Delta B_A$ are of opposite signs for $\Delta t > 2$ ps. This competition results in a negative change in $\Delta B_K$. The positive change of the demagnetization field can be easily understood, as $\Delta B_D$ is essentially the change of the $z$ component of the magnetization vector. With the elevation of temperature (c.f. inset to Fig. \ref{hk}), the magnitude of the magnetization vector is reduced and $\Delta m _z < 0$ as shown in both Figs. \ref{pma} and \ref{dm}, therefore the change of the demagnetization field is always positive. The sign change of $\Delta B_A$ is intriguing. It is a natural result of the dynamic evolution of $T_e$ and $T_p$, which is itself the driving force for the ultrafast demagnetization observed at short timescale ($\Delta t <$ 1 ps in Fig. \ref{mosig}). As can be seen from the inset to Fig. \ref{hk}, before an equilibrium is reached, the electron temperature $T_e$ is higher than the phonon temperature $T_p$. From Eq. (\ref{bk}), a higher $T_e$, whose direct consequence is a smaller $m$ ($ < m (T_p)$), will give a positive change of $B_A$, compared with the value before the arrival of the laser pulses. Once an equilibrium is established between $T_e$ and $T_p$ (Actually, in the inset to Fig. \ref{hk}, there is a small amplitude overshooting of the phonon temperature, which is solely resulted from the fact that only the heat dissipation due to electron heat conduction is considered in Eq. (\ref{mllg})), their common value is still higher than the ambient temperature, $T \approx T_e \approx T_p > T_{am}$. This results in $B_A \propto m m_z (T)$, which is smaller than its corresponding value at ambient temperature, assuming the polar angle of $\textbf{m}$ (hence $m_z$) is not increased in the whole process (Fig. \ref{hk}, $\Delta B_D$ curve, and Figs. \ref{pma} and \ref{dm}). The resulted change of anisotropy is thus negative. The above analysis qualitatively explains the change of sign for $\Delta B_K$, and hence the phase of the magnetization precession. Without the sign change in $\Delta B_K$, the magnetization precession will follow the $\Delta B_D$ curve, as plotted in Fig. \ref{pma}. Therefore, the time evolution of $\Delta B _A$ demonstrates unambiguously that the phonon temperature dependence of the interface anisotropy gives rise to the sign change in $\Delta B _K$, which is otherwise puzzling as the demagnetization field overwhelms the anisotropy field in equilibrium. The resolution of the puzzle is intricate but natural, as we know the interface anisotropy is there, although made hidden by the dominating demagnetization field in equilibrium. Dynamically, due to the phonon temperature dependence of the interface anisotropy, $\Delta B _A$ is greater than $\Delta B _D$ in magnitude. Another benefit of fathoming so deep into the connection between the ultrafast demagnetization and the following magnetization precession is that the achieved agreement between experiment and theory affords a holistic picture for laser-induced magnetization precession in ferromagnetic metal films: the driving force behind the magnetization precession is the dynamic evolution of the anisotropy field, which is directly derived from the equilibration process of the electron and phonon subsystems initiated by irradiation of ultrashort laser pulses.
\begin{figure}\centering
\begin{minipage}[c]{0.5\linewidth}
\includegraphics[width=\linewidth]{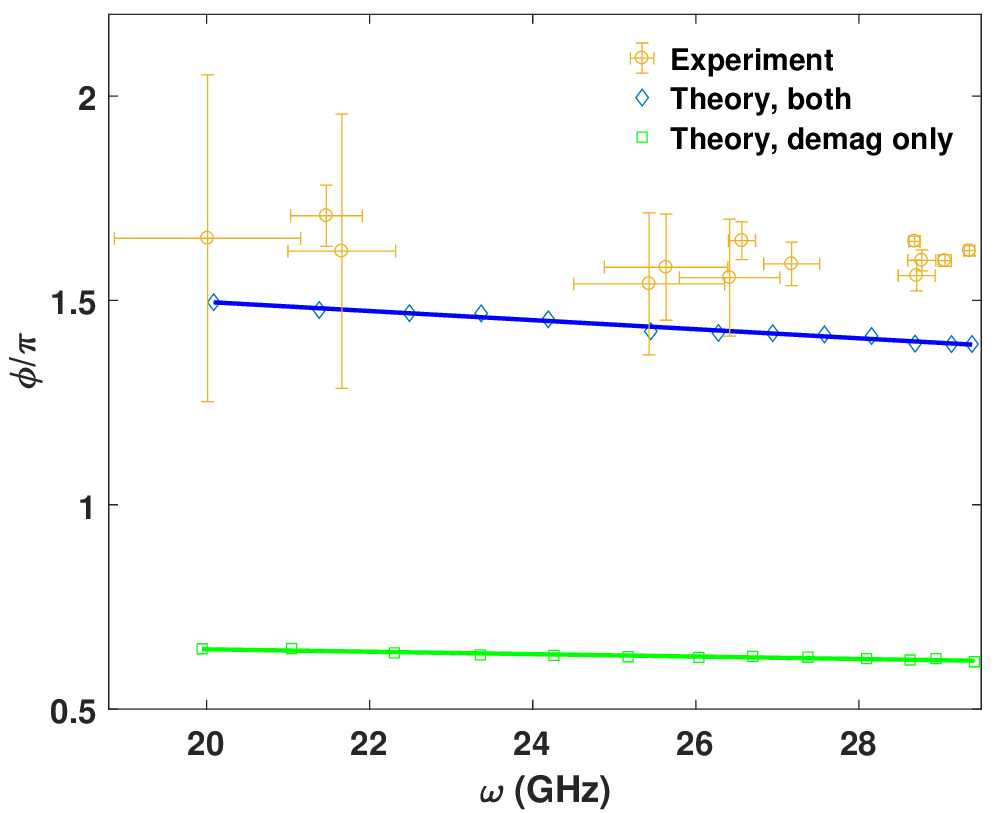}
\end{minipage}
\caption{Phase fitted from the measured data and the simulated magneto-optical traces with contribution from only $m _z$ in the presence of the interface anisotropy ($K _s \neq 0$) or not ($K _s = 0$). The lines are linear fittings to the theoretical data. For clarity, $\omega$ for the data with $K _s = 0$ is scaled down by a factor of 1.53.}
\label{phi}
\end{figure}

Finally, the actual phase of magnetization precession has to be determined to verify the claimed $\pi$ shift induced by the interface anisotropy. According to the distinct characteristics of the magnetization dynamics on both the short and long time scales, we can separate the magnetization dynamics into two stages: the ultrafast demagnetization stage, including the following recovery, and the magnetization precession stage. If we are only interested in the magnetization precession occurring on the nanosecond time scale, the effect of the ultrashort laser pulses, mediated through the elevated temperature for electrons and phonons, can be viewed as an impulse to the magnetization, similar in nature to the impulse given to a football to kick it off. Then the net effect of the laser irradiation is just to initiate the observed magnetization precession at frequency $\omega$, which is characterized by a phase $\phi _0$ at delay time $\Delta t _0$ in the form of $\sin (\omega (\Delta t - \Delta t _0) + \phi _0)$. The initial, or incubation, delay time $\Delta t _0$ is a measure of how rapid a magnetization precession is established after the irradiation of laser pulses, with $\phi _0$ being the corresponding phase. Experimentally, only the combination $\phi = \phi _0 - \omega \Delta t _0$ can be determined by fitting the measured long-term oscillation to an attenuated sine function \cite{van Kampen02prl,van Kampen02jmmm,Schellekens13} with amplitude $A _4$ and decay time $\tau _d$,
\begin{equation}
\Delta MO = A _1 + A _2 e^{-\Delta t/\tau _e} + \frac {A _3} {\sqrt{1 + \Delta t/\tau _0}} + A _4 e^{-\Delta t/\tau _d} \sin (\omega \Delta t + \phi).
\label{fitfunction}
\end{equation}
$A _1$ is a time-independent background contribution, while $A _2$ and $A _3$ correspond to the electron temperature relaxation with time constant $\tau _e$ and the one-dimensional heat diffusion to the substrate with initial time $\tau _0$, respectively. The fitted phase $\phi$ is plotted in Fig. \ref{phi}. However, the fact that the our measured signal $\Delta MO$ is related to a linear combination of all three components of the time-varying magnetization, rather than the pure $z$ component, complicates further the determination of $\phi _0$. To determine unambiguously both $\Delta t _0$ and $\phi _0$, we have performed simulations using the fitted parameters as listed in Table \ref{fitparas}, and then fit the generated oscillation with contribution from only the $z$ component of $\textbf{m}$, $m _z$, to the same fitting function, Eq. (\ref{fitfunction}). The slope and intercept of the fitted effective phase $\phi = \phi _0 - \omega \Delta t _0$ as a function of $\omega$ gives $\Delta t _0$ and $\phi _0$ separately. The results obtained using this procedure are shown in Table \ref{phi-fit}. For a direct comparison, we also list in Table \ref{phi-fit} the expected initial precession phase in adoption of the assumption that the change of the effective anisotropy can be described by a step function at time delay $\Delta t _0$. We can see immediately that the fitted phase differs from the ideal, step-like behavior expected for both the cases $K _s = 0$ and $K _s \neq 0$. Surprisingly, the phase difference $\Delta \phi$ between the fitted and the ideal values is the same for both cases, which is about 0.2$\pi$ in advance. This additional phase can be viewed as the effective phase accumulated for the magnetization precession during the ultrafast demagnetization process. As the demagnetization time is not significantly affected by the anisotropy, the accumulated phase is almost the same, irrespective of the presence of the interface anisotropy. As a result, the phase shift caused by the presence of the interface anisotropy is still a whole $\pi$ within uncertainties, although the values of the individual phases deviate from the ideal values by a finite amount and definite time-independent initial phase can only be obtained after the incubation delay time. The appearance of both $\Delta \phi$ and $\Delta t _0$ is caused by the demagnetization process, which sets the initial conditions for the subsequent magnetization precession. $\Delta t _0$ is of great importance in the quantitative description of the laser-induced magnetization precession, as it corresponds to the specific time delay after which a constant initial phase of precession can be defined; without it, the phase is frequency dependent.
\begin{table}[htb]
\begin{tabular}{|c|c|c|c|}
\hline
&$\phi _0$($\pi$)&$\phi _H$($\pi$)&$\Delta t _0$(ps)\\
\hline
$K _s$ = 0&$-$0.29 $\pm$ 0.02&$-$0.5&1.9 $\pm$ 0.4\\
\hline
$K_s \neq$ 0&0.72 $\pm$ 0.04&0.5&11 $\pm$ 2\\
\hline
\end{tabular}
\caption{Fitted initial phase $\phi _0$ and the incubation time $\Delta t _0$ in the presence ($K _s \neq$ 0) and absence ($K _s$ = 0) of the interface anisotropy. $\phi _H$ is the ideal phase for a step-like change of anisotropy.}
\label{phi-fit}
\end{table}

\section{Conclusion}
\label{conclusion}
Laser-pumped magnetization precession in Pt/Co/Pt thin film system with hidden perpendicular interface anisotropy was investigated by time resolved magneto optical Kerr effect. The measured precession can be described by a microscopic three temperature model in combination with the Landau-Lifshitz-Gilbert equation. The agreement between theory and experiment provides insight into the different roles played by the demagnetization field and the interface anisotropy field in laser-induced magnetization precession. Specifically, the phase of the precession is determined by a competition between the dynamic interface anisotropy and the demagnetization energy, which follow the phonon temperature and the electron temperature respectively. This competition results in a $\pi$ phase shift for magnetization precession in the presence of the hidden interface anisotropy, in addition to the ubiquitous demagnetization anisotropy. Due to the influence of the preceding ultrafast demagnetization process, a definite initial phase of precession can only be defined after the incubation time delay, with a value also different from the ideal value that can be expected for an instantaneous response of the anisotropy.

\begin{acknowledgments}
Supervision and guidance of Prof. Bert Koopmans on TRMOKE experiment is gratefully acknowledged. Dr. Adrianus Johannes Schellekens kindly shared his code on M3TM simulation of magnetic multilayers and critically evaluated the first draft of the manuscript.
\end{acknowledgments}

\appendix*
\section{Frequency and decay time from LLG equation}
\begin{figure}\centering
\begin{minipage}[c]{0.5\linewidth}
\includegraphics[width=\linewidth]{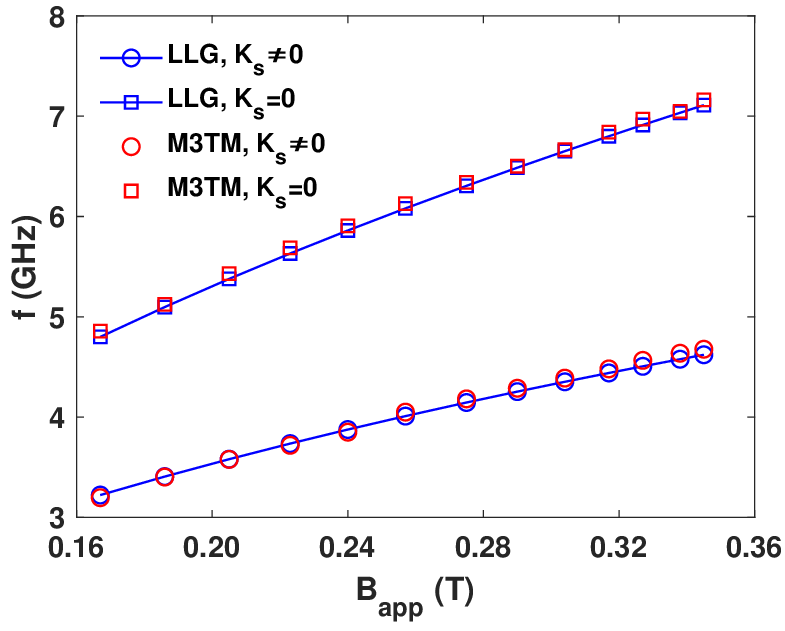}
\end{minipage}
\caption{Magnetization precession frequency $f$ predicted by the LLG equation with the interface anisotropy ($K _s \neq 0$) or not ($K _s = 0$), as compared to the simulated one from the combination of the M3TM model and the LLG equation.}
\label{freq}
\end{figure}
The oscillation frequency $\omega$ and the decay time $\tau _d$ as defined in Eq. (\ref{fitfunction}) around the equilibrium angle $\theta$ measured from the film plane for a macrospin can be obtained by linearizing the LLG equation around $\theta$, assuming that the effect of laser heating can be modelled by a step change of the magnetization direction followed by relaxation to the original equilibrium angle. The results are
\begin{equation}
\label{eqomega}
\omega = \gamma \left(\frac {B _1 B _2} {1 + \alpha ^2} - \frac {1} {\gamma ^2 \tau _d ^2}\right) ^{1/2},
\end{equation}
and
\begin{equation}
\label{eqtaud}
\tau _d = \frac {2 (\alpha + \alpha ^{-1})} {\gamma (B _1 + B _2)}.
\end{equation}
$B _2 = B _{app} \sin (\beta + \theta) - |B _K| \sin ^2 \theta$ is the magnitude of the equilibrium field, which is aligned with the equilibrium magnetization direction, and $B _1 = |B _K| \cos ^2 \theta + B _2$. $\beta$ is the angle subtended between the applied field $B _{app}$ and the $z$ direction, which is perpendicular to the film plane. The equilibrium angle $\theta$ is determined by the applied external field through the equation $B _{app} \cos (\beta + \theta) = |B _K| \sin \theta \cos \theta$. The specific forms of the precession frequency and the decay time, which is just the inverse of the imaginary part of the complex precession frequency $\tilde{\omega}$ while $\omega$ is the real part of $\tilde{\omega}$, shows an interesting feature: the relationship between the modulus of $\tilde{\omega}$ and the field is identical to the dissipationless case, provided the gyromagnetic ratio $\gamma$ is renormalized by $\sqrt {1 + \alpha ^2}$, rather than the usual renormalization factor $1 + \alpha ^2$ that can be inferred from the LL form of the LLG equation.

\begin{figure}\centering
\begin{minipage}[c]{0.5\linewidth}
\includegraphics[width=\linewidth]{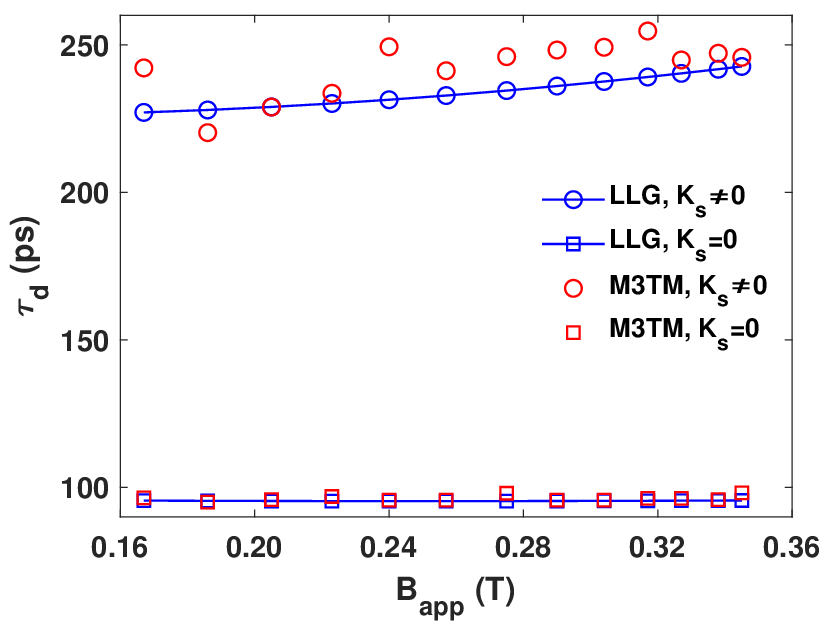}
\end{minipage}
\caption{Similar comparison as in Fig. \ref{freq} for the decay time $\tau _d$ of the magnetization precession.}
\label{taud}
\end{figure}

Corresponding to values of the externally applied field along $\beta = 0.04 \pi$, a series of $\omega$ and $\tau _d$ can be determined using Eq. (\ref{fitfunction}) from the magnetization precession produced by our M3TM-based model fed with the best-fit parameters listed in Table \ref{fitparas}, as was done in Fig. \ref{phi}. The same quantities can also be computed from Eqs. (\ref{eqomega}) and (\ref{eqtaud}). For a direct comparison, they are plotted together in Figs. \ref{freq} and \ref{taud}. For the frequency $f = \omega /2 \pi$, the LLG prediction agrees with our model result fairly well. In contrast, the agreement for $\tau _d$ between the two approaches is not that satisfactory, although there is almost no systematic variation for $\tau _d$ generated by our M3TM-based model.

\end{document}